\documentclass[a4paper]{article}
\pdfoutput=1

\usepackage{graphicx,wrapfig}
\usepackage{lipsum}
\usepackage{amsmath,amssymb,mathrsfs}
\usepackage{hyperref}
\usepackage{url}
\usepackage{mathtools}
\usepackage{changepage}
\usepackage{multirow}
\usepackage{wrapfig}
\usepackage{subfig}
\usepackage{color}
\usepackage{epsfig,enumerate,amsmath,amsfonts,amssymb,amsthm,mathrsfs,ifpdf}
\usepackage{indentfirst,relsize}
\usepackage{setspace,graphicx}
\usepackage{latexsym}
\usepackage[all]{xy}
\usepackage[usenames,dvipsnames]{pstricks}
\usepackage{pst-grad} % For gradients
\usepackage{pst-plot} % For axes
\usepackage[margin = 2.50cm]{geometry}

\usepackage{algorithm}

\usepackage[noend]{algpseudocode}

\newcommand{\remove}[1]{}

\newtheorem{theo}{Theorem}%[section]
\newtheorem{lem}[theo]{Lemma}

\newtheorem{coro}[theo]{Corollary}

\newtheorem{defi}[theo]{Definition}

\graphicspath{{./images/}}

\title{{Algorithmic study on liar's vertex-edge domination problem}}

\author{Debojyoti Bhattacharya\footnote{Inian Instittute of Technology Patna, Bihta, 801106, Bihar, India. email: debojyoti\_2021ma11@iitp.ac.in} \and Subhabrata Paul\footnote{Inian Instittute of Technology Patna, Bihta, 801106, Bihar, India. email: subhabrata@iitp.ac.in}}

\date{}
\begin{document}
\maketitle

\begin{abstract}
	Let $G=(V,E)$ be a graph. For an edge $e=xy\in E$, the closed neighbourhood of $e$, denoted by $N_G[e]$ or $N_G[xy]$, is the set $N_G[x]\cup N_G[y]$. A vertex set $L\subseteq V$ is liar's vertex-edge dominating set of a graph $G=(V,E)$ if for every $e_i\in E$, $|N_G[e_i]\cap L|\geq 2$ and for every pair of distinct edges $e_i$ and $e_j$, $|(N_G[e_i]\cup N_G[e_j])\cap L|\geq 3$.  This paper introduces the notion of liar's vertex-edge domination which arises naturally from some applications in communication networks. Given a graph $G$, the \textsc{Minimum Liar's Vertex-Edge Domination Problem} (\textsc{MinLVEDP}) asks to find a liar's vertex-edge dominating set of $G$ of minimum cardinality. In this paper, we study this problem from algorithmic point of view. We show that \textsc{MinLVEDP} can be solved in linear time for trees, whereas the decision version of this problem is NP-complete for chordal graphs, bipartite graphs, and $p$-claw free graphs for $p\geq 4$. We further study approximation algorithms for this problem. We propose two approximation algorithms for \textsc{MinLVEDP} in general graphs and $p$-claw free graphs.
	%We propose an $O(\ln \Delta(G))$-approximation algorithm for \textsc{MinLVEDP} in general graphs, where $\Delta(G)$ is the maximum degree of the input graph. Also, we design a constant factor approximation algorithm for $p$-claw free graphs.
	On the negative side, we show that the \textsc{MinLVEDP} cannot be approximated within $\frac{1}{2}(\frac{1}{8}-\epsilon)\ln|V|$ for any $\epsilon >0$, unless $NP\subseteq DTIME(|V|^{O(\log(\log|V|)})$. Finally, we prove that the \textsc{MinLVEDP} is APX-complete for bounded degree graphs and $p$-claw free graphs for $p\geq 6$. 
	
	%Insert your abstract here. Include keywords, PACS and mathematical
	%subject classification numbers as needed.
\noindent\textbf{keywords:} {Liar's vertex-edge dominating set, NP-completeness, Approximation algorithms, Chordal graphs, Bipartite graphs}
%	\subclass{ 05C85 \and  05C05 \and 05C69}
\end{abstract}

\section{Introduction}
\label{intro}

Let $G=(V,E)$ be a graph with vertex set $V$ and edge set $E$. For a vertex $v\in V$, the sets $N_G(v)=\{u\in V|uv\in E\}$ and $N_G[v]=N_G(v)\cup \{v\}$ are called open neighbourhood and closed neighbourhood of $v$, respectively. We denote the degree of a vertex $v\in V$ by $deg_G(v)$, which is defined as $deg_G(v)=|N_G(v)|$. For an edge $e=xy\in E$, the closed neighbourhood of $e$, denoted by $N_G[e]$ or $N_G[xy]$, is the set $N_G[x]\cup N_G[y]$. An edge $e$ is said to be \emph{vertex-edge dominated} (or \emph{ve-dominated}) by a vertex $v$ if $v\in N_G[e]$. A subset $D\subseteq V$ is called a \emph{vertex-edge dominating set} (or a \emph{ve-dominating set}) of $G$ if for every edge $e\in E$, $|N_G[e]\cap D|\geq 1$. The cardinality of a minimum ve-dominating set is called the \emph{ve-domination number} of $G$ and it is denoted by $\gamma_{ve}(G)$. Peters introduced ve-domination in his Ph.D. thesis in 1986 \cite{peters}. But it did not get much attention until Lewis proved some interesting results \cite{lewis}. Since then, ve-domination has been extensively studied in literature \cite{boutrig2016vertex,jena,paul2,zylinski2019vertex}. For some integer $k$, a subset $D_k\subseteq V$ is called a \emph{$k$-ve dominating set} of $G$ if $|N_G(e)\cap D_k|\geq k$ for every edge $e\in E$. The cardinality of a minimum $k$-ve dominating set is called the \emph{$k$-ve domination number} of $G$ and it is denoted by $\gamma_{kve}(G)$. This variation is also well studied in literature \cite{krishna,li2023polynomial,naresh}. Different variations of ve-domination, namely independent ve-domination \cite{lewis,paul}, total ve-domination \cite{ahangar2021total,Totalve-domchellali}, global ve-domination \cite{chitra2012global,globalvedom} etc. We recall one more variation of the classical domination problem, namely \emph{liar's domination problem}. A subset of $D\subseteq V$ is called liar's dominating set if $|N_G[v_i]\cap D|\geq 2$, for every $v_i\in V$ and $|(N_G[v_i]\cup N_G[v_j])\cap D|\geq 3$ for every pair of distinct vertices $v_i,v_j\in V$. The liar's domination problem is also well studied in literature \cite{unitdiscliar,PANDAliarscomplexity,panda2015hardness,roden2008liar,slater2009liar}. In this article, we introduce a variation of ve-domination problem, namely liar's vertex-edge domination problem, which arises naturally from the following application.  

Let us consider a communication network, represented by the graph $G=(V,E)$, where an edge $e=xy\in E$ represents a communication link between its endpoints $x$ and $y$. For the smooth functioning of this communication network, we need to monitor every link constantly so that every broken or damaged link can be identified immediately. This monitoring can be done by installing sentinels at the vertices. The job of a sentinel placed at a vertex $v$ is in two steps; $(i)$ to detect the presence of a broken or damaged link which is incident to any vertex of $N_G[v]$ and then $(ii)$ to report that broken or damaged link. In other words, every link $e=xy\in E$ is monitored by the sentinels that are present at $N_G[x]\cup N_G[y]$. We need to keep in mind that to make this monitoring cost-effective, we need to install sentinels at the minimum number of vertices as each sentinel can be of high cost. Therefore, the optimization problem in this scenario is to find the minimum set of vertices at which the sentinels are to be placed. Note that a minimum ve-dominating set of $G$ is a solution to that optimization problem. In some cases, we want to make the network fault-tolerant, that is, any broken or damaged link in the network must be detected even if a few sentinels have failed to detect the broken or damaged edge. In that case, each link needs to be monitored by more than one sentinel. If we assume that in $N_G[e]$, at most $k-1$ ($k\geq 2$) sentinels are faulty, then 
by placing the sentinels at a minimum $k$-ve dominating set of $G$, we can secure the network.

In some undesired situation, it might happen that a sentinel placed at $v$ correctly detects the broken or damaged link incident to a vertex of $N_G[v]$ but misreports an undamaged link incident to a vertex of $N_G[v]$ as a damaged channel. Let us assume that every sentinel in the neighbourhood of a broken or damaged edge $e$ detects $e$ as a damaged edge but at most one sentinel at $v$ in $N_G[e]$ is misreporting an undamaged link incident to a vertex of $N_G[v]$ as a damaged channel. In this case, clearly, a $2$-ve dominating set is not sufficient. Under this assumption, to secure the network we need to install the sentinels at a set of vertices that has some special property. We define this set of vertices as a \emph{liar's vertex-edge dominating set} (or liar's ve-dominating set) of $G$.

%\begin{definition}
%Given a graph $G=(V,E)$,  let $L$ be a subset of vertices of $G$ having the following property: 
% 
%If any damaged or broken channel $e\in E$, every sentinel placed at the set of vertices in $N_G[e]\cap L$ except one correctly reports $e$ as broken and at most one sentinel $y$ in $N_G[e]\cap L$ reports some other edge $e'$ incident to the vertices of $N_G[y]$ or does not report any edge. 
%
%\noindent If $e$ is correctly identified by $L$ as a broken channel, then $L$ is a liar's ve-dominating set of $G$. 
%\end{definition}

First, let us discuss a few properties of such a liar's ve-dominating set of a graph. If $L\subseteq V$ is a liar's ve-dominating set of $G$, then clearly $N_G[e]\cap L$ must contain at least $2$ vertices. Otherwise, if the edge $e$ is damaged and there is one sentinel in $N_G[e]$ and above all, that sentinel misreports some other edge as damaged edge, then it is impossible to detect the actual damaged edge $e$. On the other hand, if $D\subseteq V$ is a $3$-ve dominating set of $G$, then $D$ is also a liar's ve-dominating set. Because in that case, each edge $e$ has at least $3$ sentinels in $N_G[e]$ and by our assumption at most one sentinel can misreport. Therefore, two other sentinels correctly identify the edge $e$ if it is broken. 
Hence, every liar's ve-dominating set is a $2$-ve dominating set and every $3$-ve dominating set is a liar's ve-dominating set. The following lemma characterizes a liar's dominating set of a graph. 

%Clearly, for a connected graph $G$ having at least $3$ vertices, $\gamma_{2ve}(G)\leq\gamma_{lve}(G)\leq\gamma_{3ve}(G)$. 

\begin{lem}\label{Lem:Definitionlved}
	Let $G=(V,E)$ be a graph. A vertex set $L\subseteq V$ is liar's ve-dominating set if and only if it satisfies the following conditions:
	\begin{itemize}
		\item[(i)] for every $e_i\in E$, $|N_G[e_i]\cap L|\geq 2$ and
		\item[(ii)] for every pair of distinct edges $e_i$ and $e_j$, $|(N_G[e_i]\cup N_G[e_j])\cap L|\geq 3$. 
	\end{itemize}
\end{lem}

\begin{proof}
	Let $L$ be a liar's ve-dominating set of $G$. From the previous discussion, it is clear that  $|N_G[e_i]\cap L|\geq 2$, for every $e_i\in E$. Also, for some damaged edge $e_i$ if one out of two sentinels at $N_G[e_i]$ reports some other edge $e_j$ and the other one reports $e_i$, then to correctly identify the damaged edge $e_i$, there must be another sentinel either at $N_G[e_i]$ or at $N_G[e_j]$. This extra sentinel correctly reports the damaged edge. Thus, $|(N_G[e_i]\cup N_G[e_j])\cap L|\geq 3$, for every pair of distinct edges $e_i$ and $e_j$.
	
	% \ref{Obs:N[e]intersectionLgeq2}, $L$ is a double vertex-edge dominating set of $G$. Therefore, $|N[e]\cap L|\geq 2$. Suppose that there exists a pair of edges $e_1$ and $e_2$ such that $|(N[e_1]\cup N[e_2])\cap L|<3$. Since every edge is double vertex-edge dominated, we have $|(N[e_1]\cup N[e_2])\cap L|=2$. Let $\{x,y\}\in (N[e_1]\cup N[e_2])\cap L$. Suppose that the tracker at $x$ reports the edge $e_1$ and the tracker at $y$ reports $e_2$. Now, one of them can lie and thus $L$ can not detect the defective edge which is a contradiction. Therefore, $|(N[e_1]\cup N[e_2])\cap L|\geq 3$.	
	Conversely, let $L$ be a subset of $V$ satisfying the conditions $(i)$ and $(ii)$. We place the sentinels at the vertices of $L$. Since $L$ satisfies condition $(i)$, if $e_i$ is a damaged edge, then there are at least two sentinels at $N_G[e_i]$. This implies that there is always one sentinel that identifies a damaged edge correctly. Also, if a sentinel at $N_G[e_i]$ misreports some other edge $e_j$ as damaged edge, then the third sentinel at $N_G[e_i]\cup N_G[e_j]$ correctly report the damaged edge. Therefore, the damaged edge is reported by at least two sentinels. By our assumption, it is clear that if any edge is reported by two sentinels as a damaged edge, it must be an actual damaged edge. Therefore, any damaged edge in $E$ can be correctly recognized by $L$. Hence, $L$ is a liar's ve-dominating set. \qed
\end{proof}
%
%the other one in $N[e]$ correctly identifies the $e$. Now, if among the two trackers of $e$, one reports $e$ and other reports some other edge $e'$, then by condition (b), there is a vertex in $N[e]\cup N[e']$ which correctly identifies the defective edge. Thus, $L$ is a liar's vertex-edge dominating set.  

The above Lemma \ref{Lem:Definitionlved} gives a definition of a liar's ve-dominating set in a graph $G$. We are going to use this characterization as our definition. 

\begin{defi}
	A vertex set $L\subseteq V$ is \emph{liar's ve-dominating set} of a graph $G=(V,E)$ if 
	\begin{itemize}
		\item[(i)] for every $e_i\in E$, $|N_G[e_i]\cap L|\geq 2$ and
		\item[(ii)] for every pair of distinct edges $e_i$ and $e_j$, $|(N_G[e_i]\cup N_G[e_j])\cap L|\geq 3$. 
	\end{itemize}
	The minimum cardinality of a liar's ve-dominating set of $G$ is called the \emph{liar's ve-domination number} and it is denoted by $\gamma_{lve}(G)$. 
\end{defi}
In this paper, we have studied the algorithmic status of liar's ve-domination problem. The minimum liar's ve-domination problem and its corresponding decision version are described as follows:

\noindent\underline{\textsc{Minimum Liar's VE-Domination Problem} (\textsc{MinLVEDP})}

\noindent\emph{Instance}: A graph $G=(V,E)$.

\noindent\emph{Output}: A minimum liar's ve-dominating set $L_{ve}$ of $G$.

%\noindent\emph{Measure}: Cardinality of the liar's ve-dominating set.

\noindent\underline{\textsc{Liar's VE-Domination Decision Problem} (\textsc{DecideLVEDP})}

\noindent\emph{Instance}: A graph $G=(V,E)$ and an integer $k$.

\noindent\emph{Question}: Does there exist a liar's ve-dominating set of size at most $k$?

%A graph $G=(V,E)$ is $p$-claw free graph if, for every vertex $v\in V$, the induced subgraph of $N_G(v)$ does not contain a independent set of size $p$.

The results presented in this paper are summarized as follows:
\begin{itemize}
	\item[(i)] We design a linear time algorithm for \textsc{MinLVEDP} in trees.
	\item[(ii)] We prove that \textsc{DecideLVEDP} is NP-complete for general graphs, chordal graphs, bipartite graphs and $p$-claw free graphs for $p\geq 4$.
	\item[(iii)] We also design an $O(\ln \Delta(G))$-approximation algorithm for \textsc{MinLVEDP}, where $\Delta(G)$ is the maximum degree of the input graphs.
	\item[(iv)] We propose a $3(p-1)$-approximation algorithm for \textsc{MinLVEDP} in $p$-claw free graphs. 
	\item[(v)] We prove that the \textsc{MinLVEDP} cannot be approximated within $\frac{1}{2}(\frac{1}{8}-\epsilon)\ln|V|$ for any $\epsilon >0$, unless $NP\subseteq DTIME(|V|^{O(\log(\log|V|)})$.
	\item[(vi)] We show that the \textsc{MinLVEDP} is APX-complete for bounded degree graphs and $p$-claw free graphs for $P\geq 6$.
\end{itemize}

The rest of the paper is organized as follows. In Section $2$, we propose a linear time algorithm for \textsc{MinLVEDP} in trees. In Section $3$, we prove the NP-completeness result for \textsc{DecideLVEDP} in general graphs, chordal graphs, bipartite graphs, and $p$-claw free graphs for $p\geq 4$. In section $4$, we first design approximation algorithms for \textsc{MinLVEDP} for general graphs and $p$-claw free graphs. We then show a hardness of approximation result for \textsc{MinLVEDP} for general graphs. After that, we prove that \textsc{MinLVEDP} is APX-complete for bounded degree graphs and $p$-claw free graphs for $p\geq 6$. Finally, Section $5$ concludes the paper with some directions for further study.

\section{Algorithm in Tree}
In this section, we propose a linear time algorithm to find the minimum liar's ve-dominating set of a given tree. Our proposed algorithm is a labelling based greedy algorithm. First, we define a generalization of liar's ve-dominating set, namely $l(k,t)$-dominating set of an $l$-labelled graph. A graph $G=(V,E)$ is called an $l$-labelled graph if every edge $e$ is labelled as $k(e)$ where $k(e)\in \{0,1,2\}$ and every vertex $v$ is labelled as $t(v)$, where $t(v)\in \{B,R\}$. A subset $D$ of $V$ is called $l(k,t)$-dominating set of $G$ if and only if it satisfies the following conditions:
\begin{enumerate}
	\item[(i)] if $t(v)=R$, then $v\in D$,
	\item[(ii)] for every edge $e$, $|N_{G}[e]\cap D|\geq k(e)$ and
	\item[(iii)] for every pair of distinct edges $e$ and $e'$, $|(N_{G}[e]\cup N_{G}[e'])\cap D|\geq k(e)+k(e')-1$.
\end{enumerate}
Note that if $t(v)=B$ for every vertex $v\in V$ and $k(e)=2$ for every edge $e\in E$, then a minimum $l(k,t)$-dominating set is a minimum liar's ve-dominating set. Next, we describe an algorithm to compute an $l(k,t)$-dominating set of a given tree $T$ where the labels are initialized as $k(e)=2$ for all $e\in E$ and $t(v)=B$ for all $v\in V$.

\begin{algorithm}
	\caption{\small $l(k,t)$-DS\_Tree$(T,l(k,t))$}
	\textbf{Input:} A labelled tree $T=(V,E)$ with $k(e)=2$ for all $e\in E$ and $t(v)=B$ for all $v\in V$\\
	\textbf{Output:} A minimum $l(k,t)$-dominating set $L$ of $T$.
	\label{Algo:treemain}
	\begin{algorithmic}[1]
		\State \small Find a ordering $\sigma=\{v_1,v_2,\ldots,v_n\}$ such that $\{v_n,v_{n-1},\ldots,v_1\}$ is a BFS ordering;

		\State \small$L=\emptyset$, $d(v)=deg_T(v)$ and $r(v)=\emptyset$ for all $v\in V$;
		\For{$(i=(h-1)~ to~ 1)$} ~~~~~~~~~~~~~~~~~~~~~~~~~~~~~[$h$ be the height~of~$T$ induced by $\sigma$]
		\For{$(every~ support~ vertex~ u\in \sigma~ at~ i)$}
		%\State Compute $d(u)$, $r(u)$ and $r(p(u))$;
		%\If{$(u=root)$}
		%		\State $D=D\cup r(u)$.
		%		\If{$(d_u\geq 2)$}
		%		\State include $3-|r(u)|$ vertices of $N_{T}[u]$ in $D$ with priority $u$.
		%		\ElsIf{$(d_u=1)$}
		%		\State include $2-|r(u)|$ vertices of $N_{T}[u]$ in $D$ with priority $u$.
		%		\Else
		%		\If{$(\exists~ an~ edge~ e~incident~to~u~ with~ k(e)=1)$}
		%		\State include $1-|r(u)|$ vertices of $N_{T}[u]$ in $D$ with priority $u$.
		%		\EndIf
		%		\EndIf
		%		\State $T=T\setminus N_{T}[u]$.
		%\Else	
		\If{$(d(u)\geq 3)$}
		\State Mark $(3-|r(u)|)$ many maximum neighbours of $u$ in $N_T[u]\setminus r(u)$ as $R$;
		%\State Relabel $3-|r(u)|$ vertices of $N_{T}[u]$ as $R$ with priority $(w,u,v)$;
		
		\ElsIf{$(d(u)=2)$}
		\State Mark $(2-|r(u)|)$ many maximum neighbours of $u$ in $N_T[u]\setminus r(u)$ as $R$;
		%\State Relabel $2-|r(u)|$ vertices of $N_{T}[u]$ as $R$ with priority $(w,u)$;
		
		\ElsIf{$(there~is~ an~ edge~ e~incident~to~u~with~ k(e)=1)$}
		\State Mark $(1-|r(u)|)$ many maximum neighbours of $u$ in $N_T[u]\setminus r(u)$ as $R$;
		%\State Relabel $1-|r(u)|$ vertices of $N_{T}[u]$ as $R$ with priority $w$;
		
		\EndIf
		\State Update $r(v)$ of all $v\in N_T[v_r]$, where $v_r$ is the newly marked vertex;
		%\EndIf
		\EndFor		
		%		\If{$(w\neq~root)$}
		%		\For{$(every ~u\in c(w))$}
		%		\If{$(d_u=2~\&~|r(u)\cup r(w)|<3)$}
		%		\State Relabel $t(p(w))=R$
		%		\EndIf
		%		\State $D=D\cup (c(u)\cap r(u))$
		%		\State $T=T\setminus c(u)$
		%		\State $k(uw)=\max\{k(uw)-|c(u)\cap r(u)|,0\}$
		%		\EndFor
		%		\Else
		%\If{$(d_u=2~\&~|r(u)\cup r(w)|<3)$}
		%		\If{$(\exists~u\in N_{T}[p(u)]~such~that~d(u)=2~\&~|r(u)\cup r(p(u))|<3)$}
		%		\State Mark the maximum neighbour of $u$ as $R$
		%\State Relabel a vertex $N_T[w]$ as $R$ with priority $p(w)$;
		%\EndIf
		\For{$(every~ support~ vertex~ u\in \sigma~ at~ i)$}
		\If{$(d(u)=2~\&~|r(u)\cup r(p(u))|<3)$}
		\State Mark the maximum neighbour of $p(u)$ in $N_T[p(u)]\setminus r(p(u))$ as $R$;
		\State Update $r(v)$ of all $v\in N_T[v_r]$, where $v_r$ is the newly marked vertex;
		\EndIf
		\State $L=L\cup (c(u)\cap r(u))$;
		\State $T=T\setminus c(u)$;
		\State Update $r(u)$;
		\State Update $k(u p(u))$ as $k(u p(u))=\max\{k(up(u))-|c(u)\cap r(u)|,0\}$;
		\State Update $d(p(u))$;
		\EndFor
		%\EndIf
		
		\EndFor
		\State $L=L~ \cup $~Star$(T,l(k,t))$;
		\State \Return $L$;
	\end{algorithmic}
\end{algorithm}

Next, we give a brief description of our proposed algorithm. For this, first, we state a few notations that are used in the algorithm. For a vertex $u$, let $c(u)$,  $r(u)$ and $d(u)$ denote the set of children of $u$, the set of required vertices in $N_{T}[u]$ and the number of edges incident to $u$ with $k(e)=2$, respectively. A vertex is called a support vertex if it is adjacent to a pendant vertex. Also, $p(u)$ denotes the parent of $u$. A vertex $w$ is the maximum neighbour of $u$ if $w\in N_T[u]$ is the neighbour of $u$ with maximum index with respect to $\sigma$.

To compute an $l(k,t)$-dominating set of $T$, we process the vertices according to an ordering $\sigma=\{v_1,v_2,\ldots,v_n\}$ such that $\{v_n,v_{n-1},\ldots, v_1\}$ is a BFS ordering with $v_n$ being the root. We start with an empty $l(k,t)$-dominating set $L$. We process the vertices according to the decreasing order of the levels induced by $\sigma$. At level $i$, we process only the support vertices in two rounds. Note that every vertex in that level is either a support vertex or a leaf as $\sigma$ is a reverse of BFS ordering. In the first round (lines $4$--$11$), while 
processing a support vertex $u$, based on the value of $d(u)$ and existence of a pendant edge $e$ incident on $u$ with $k(e)=1$, we mark few vertices as $R$. As a result, every edge incident to vertices of $c(u)$ and every pair of edges incident to vertices of $c(u)$ satisfy the second and third condition of the definition of $l(k,t)$-dominating set. In the second round (lines $12$--$20$), we again process the support vertices at level $i$. Under certain condition, we mark the maximum neighbour of $p(u)$ in the ordering $\sigma$ as $R$. This ensures that every pendant edge incident to $u$ satisfies the third condition of $l(k,t)$-dominating set. Then we select all the vertices from $c(u)$ that are marked as $R$ into $L$ and delete $c(u)$ from $T$. After processing all support vertices at level $i$, we process all the support vertices of the level $(i-1)$ in a similar way till we have a tree with $2$ levels which is a star. We then compute $l(k,t)$-dominating set of a star directly by calling the function Star()(see Algorithm \ref{Algo:LVEDSuroot}) and include those vertices in $L$. The algorithm is described in Algorithm \ref{Algo:treemain}.

\begin{algorithm}[H]
	\caption{\small Star$(T,l(k,t))$}
	\textbf{Input:} An $l$-labelled Star $T=(V,E)$ \\
	\textbf{Output:} A minimum $l(k,t)$-dominating set $D$ of $T$.
	\label{Algo:LVEDSuroot}
	\begin{algorithmic}[1]
		\State $D=\emptyset$;
		\If{$(d(u)\geq 2)$}
		\State Include $(3-|r(u)|)$ many vertices of $N_T[u]\setminus r(u)$ in $D$;
		\ElsIf{$(d(u)=1)$}
		\State Include $(2-|r(u)|)$ many vertices of $N_T[u]\setminus r(u)$ in $D$;
		\ElsIf{$(\exists~ an~ edge~ e~incident~to~u~with~ k(e)=1)$}
		\State Include $(1-|r(u)|)$ many vertices of $N_T[u]\setminus r(u)$ in $D$;
		\EndIf
		
		%		\If{$(d_u\geq 2)$}
		%		\State include $3-|r(u)|$ vertices of $N_{T}[u]$ in $D$ with priority $u$;
		%		\ElsIf{$(d_u=1)$}
		%		\State include $2-|r(u)|$ vertices of $N_{T}[u]$ in $D$ with priority $u$;
		%		\Else
		%		\If{$(\exists~ an~ edge~ e~incident~to~u~ with~ k(e)=1)$}
		%		\State include $1-|r(u)|$ vertices of $N_{T}[u]$ in $D$ with priority $u$;
		%		\EndIf
		%		\EndIf
		\State $D=D\cup r(u)$;
		%\State $T=T\setminus N_{T}[u]$;
		\State \Return $D$;
	\end{algorithmic}
\end{algorithm}

\subsection{Correctness of Algorithm \ref{Algo:treemain}}
In this subsection, we prove that our proposed algorithm correctly finds a minimum $l(k,t)$-dominating set of a given tree $T$. While processing a support vertex $u$ at level $i$ in the first round, we mark few vertices as $R$. In the following lemma, we show that the $l(k,t)$-domination number remains the same after relabelling.

\begin{lem}\label{lem: du geq 3}
	Let $u$ be the support vertex at level $i$ processed in the first round. 
	
	\begin{itemize}
		\item[(a)] If $d(u)\geq 3$, then $\gamma_{l(k,t)}(T)=\gamma_{l(k',t')}(T')$, where $T'$ is obtained by relabelling $3-|r(u)|$ many maximum neighbours of $u$ in $N_T[u]\setminus r(u)$ as $R$ and all the other labels remain the same.
		
		\item[(b)] If $d(u)=2$, then $\gamma_{l(k,t)}(T)=\gamma_{l(k',t')}(T')$, where $T'$ is obtained by relabelling $2-|r(u)|$ many maximum neighbours of $u$ in $N_T[u]\setminus r(u)$ as $R$ and all the other labels remain the same.
		
		\item[(c)] If $d(u)= 1$ and there exists an edge $e$ incident to $u$ with $k(e)=1$, then
		$\gamma_{l(k,t)}(T)=\gamma_{l(k',t')}(T')$, where $T'$ is obtained by relabelling $1-|r(u)|$ many maximum neighbours of $u$ in $N_T[u]\setminus r(u)$ as $R$ and all the other labels remain the same.
	\end{itemize}
\end{lem}
\begin{proof}
	(a)	Let $L$ be a minimum $l(k,t)$-dominating set of $T$. Note that at any stage of the algorithm $k(up(u))=2$. Since $d(u) \geq 3$, there are two pendant edges $e_i$ and $e_j$ incident to $u$ such that $k(e_i)=k(e_j)=2$. Therefore, $|(N_{T}[e_i]\cup N_{T}[e_j])\cap L|\geq k(e_i)+k(e_j)-1$ which implies that $|N_{T}[u]\cap L|\geq 3$. Hence, $(N_{T}[u]\cap L)$ contains a set $Y$ of size $3-|r(u)|$ such that every vertex of $Y$ is labelled as $B$. Let $X$ be the set of $3-|r(u)|$ many maximum neighbours of $N_T[u]\setminus r(u)$ that has been relabelled as $R$ in $T'$. Let $L'=(L\setminus Y)\cup X$. It can be shown that $L'$ is a $l(k',t')$-dominating set of $T'$. Therefore, $\gamma_{l(k',t')}(T')\leq \gamma_{l(k,t)}(T)$.
	%
	%, then
	%there is a vertex, say $v_i$, of $(c(u)\cup u)\setminus r(u)$ is in $D$. Let $v_m$ be the maximum neighbour of $N_T[u]\setminus r(u)$. Consider the set $D'=(D\setminus \{v_i\})\cup \{v_m\}$. Clearly, $D'$ is a $l(k',t')$-dominating set of $T'$. Finally if $3-|r(u)|=2$, then there is a pair of vertices $v_i,v_j\in (D\setminus r(u))\cap (c(u)\cup \{u\})$. Without loss of generality, suppose that $i<j$. Consider the set $D'=(D\setminus \{v_i\})\cup \{p(u)\}$ 
	%
	%Without loss of generality let us assume that $w\notin D$ and $p\neq v$. Let $D'=(D\setminus\{p\})\cap \{w\}$. Similarly, we can show that $u,v\in D'$. Clearly, $D'$ is a $L(k,t')$-dominating set of $T'$. If $p=v$ and $w,u\in D$, then we are done. Otherwise, we proceed the same way as above by replacing the vertices in $c(u)$ with $\{w,u\}$. 
	
	Conversely, let $L'$ be a minimum $l(k',t')$-dominating set of $T'$. Since $k(e)=k'(e)$ for every edge $e$ in $T$ and the set of required vertices in $T$ is a subset of the set of required vertices in $T'$. Therefore, a $l(k',t')$-dominating set of $T'$ is also a $l(k,t)$-dominating set of $T$. Hence, $\gamma_{l(k,t)}(T) \leq \gamma_{l(k',t')}(T')$. Therefore, $\gamma_{l(k,t)}(T)=\gamma_{l(k',t')}(T')$.

	%	Proof of part (b) and (c) of Lemma \ref{lem: du geq 3}.
	%\begin{proof}
	(b) Let $L$ be a minimum $l(k,t)$-dominating set of $T$. Since $d(u) = 2$, there is a pendant edges $e_i$ incident to $u$ such that $k(e_i)=2$. Therefore, $|N_{T}[e_i]\cap L|\geq k(e_i)$ which implies that $|N_{T}[u]\cap L|\geq 2$. Hence, $(N_{T}[u]\cap L)$ contains a set $Y$ of size $2-|r(u)|$ such that every vertex of $Y$ is labelled as $B$. Let $X$ be the set of $2-|r(u)|$ many maximum neighbours of $N_T[u]\setminus r(u)$ that has been relabelled as $R$ in $T'$. Let $L'=(L\setminus Y)\cup X$. It can be shown that $L'$ is a $l(k',t')$-dominating set of $T'$. Therefore, $\gamma_{l(k',t')}(T')\leq \gamma_{l(k,t)}(T)$.
	
	Conversely, let $L'$ be a minimum $l(k',t')$-dominating set of $T'$. Since $k(e)=k'(e)$ for every edge $e$ in $T$ and the set of required vertices in $T$ is a subset of the set of required vertices in $T'$. Therefore, a $l(k',t')$-dominating set of $T'$ is also a $l(k,t)$-dominating set of $T$. Hence, $\gamma_{l(k,t)}(T) \leq \gamma_{l(k',t')}(T')$. Therefore, $\gamma_{l(k,t)}(T)=\gamma_{l(k',t')}(T')$. 
	
	(c)  Let $L$ be a minimum $l(k,t)$-dominating set of $T$. Since $d(u) = 1$ and there is a pendant edges $e_i$ incident to $u$ such that $k(e_i)=1$, we have $|N_{T}[e_i]\cap L|\geq k(e_i)$ which implies that $|N_{T}[u]\cap L|\geq 1$. Hence, $(N_{T}[u]\cap L)$ contains a set $Y$ of size $1-|r(u)|$ such that every vertex of $Y$ is labelled as $B$. Let $X$ be the set of $1-|r(u)|$ many maximum neighbours of $N_T[u]\setminus r(u)$ that has been relabelled as $R$ in $T'$. Let $L'=(L\setminus Y)\cup X$. It can be shown that $L'$ is a $l(k',t')$-dominating set of $T'$. Therefore, $\gamma_{l(k',t')}(T')\leq \gamma_{l(k,t)}(T)$.
	
	Conversely, let $L'$ be a minimum $l(k',t')$-dominating set of $T'$. Since $k(e)=k'(e)$ for every edge $e$ in $T$ and the set of required vertices in $T$ is a subset of the set of required vertices in $T'$. Therefore, a $l(k',t')$-dominating set of $T'$ is also a $l(k,t)$-dominating set of $T$. Hence, $\gamma_{l(k,t)}(T) \leq \gamma_{l(k',t')}(T')$. Therefore, $\gamma_{l(k,t)}(T)=\gamma_{l(k',t')}(T')$. \qed
	%\end{proof}
	%ar to Let $D$ be a minimum $L(k,t)$-dominating set of $T$. Observe that, $k(uw)=2$. Since $d_u = 2$, there is a pendent edge $e_1$ incident to $u$ such that $k(e_1)=2$. Therefore, $|N_{T}[u]\cap D|\geq 2$. Clearly, $r(u)\subset D$. If $2-|r(u)|=0$ then we are done. Otherwise, $\{w,u\}\subset D$, then we are done. Otherwise, without loss of generality let us assume that $w\notin D$. Thus, there exist a vertex $p\in c(u)\cap D$ such that $p\neq u$. Let $D'=(D\setminus\{p\}\cap w)$. Similarly, we can show that $u\in D'$. Clearly, $D'$ is a $L(k,t')$-dominating set of $T'$.
	%
	%Let $D'$ be a minimum $L(k,t')$-dominating set of $T'$. It is easy to see that $D'$ is a $L(k,t)$-dominating set of $T$. Therefore, $\gamma_{(k,t)}(T)=\gamma_{(k,t')}(T')$.  
	%
	%
	%(c) 	Let $D$ be a minimum $L(k,t)$-dominating set of $T$. Clearly, $r(u)\subset D$ and for every edge $e$, $|N_{T}[e]\cap D|\geq k(e)$. Since $d_u=1$, we have $k(uw)=2$. Let $e_1$ be a pendent edge incident to $u$ such that $k(e_1)=1$. If $1-|r(u)|=0$, then we are done. Otherwise, there exists a vertex $p\in N_{T}[u]\cap D$ such that $t(p)=B$. Let $D'=(D\setminus \{p\})\cup \{w\}$. It is easy to see that $D'$ is a $L(k,t')$-dominating set of $T'$.
	%
	%The other direction is straightforward. Hence, $\gamma_{(k,t)}(T)=\gamma_{(k,t')}(T')$.
\end{proof}

%\begin{lemma}
%	Let $v$ be a leaf of $T$ and $u$ is parent of $v$. If $d_u= 2$, then $\gamma_{(k,t)}(T)=\gamma_{(k,t')}(T')$ where $T'$ is obtained by relabelling $2-|r(u)|$ vertices of $N_{T}[u]$ as $R$.
%\end{lemma}
%\begin{proof}
%	%	Proof is similar to Lemma \ref{lem:u not root and du geq 3}.
%
%\qed\end{proof}
%
%\begin{lemma}
%	Let $v$ be a leaf of $T$ and $u$ is parent of $v$. If $d_u= 1$ and there exists an edge $e$ incident to $u$ with $k(e)=1$, then $\gamma_{(k,t)}(T)=\gamma_{(k,t')}(T')$ where $T'$ is obtained by relabelling $1-|r(u)|$ vertices of $N_{T}[u]$ as $R$.
%\end{lemma}
%\begin{proof}
%
%\qed\end{proof}

The next two lemmas show the correctness of the algorithm for the second round.

\begin{lem}\label{lem:d(u)=2}
Let $u$ be the support vertex at level $i$ processed in the second round. Also, assume that the labels of $T$ are such that $k(e)\leq |r(u)|$ for every edge $e$ incident to a vertex of $c(u)$ and $k(e)+k(e')-1\leq |r(u)|$ for every pair of distinct edges $e$ and $e'$ incident to vertices of $c(u)$. If $d(u)=2$ and $|r(u)\cup r(p(u))|<3$, then $\gamma_{l(k,t)}(T)=\gamma_{l(k',t')}(T')+|c(u)\cap r(u)|$, where $T'$ is obtained from $T$ by deleting the vertex set $c(u)$, relabelling the maximum neighbour of $p(u)$ in $N_T[p(u)]\setminus r(p(u))$ as $R$, setting  $k'(up(u))=\max\{(k(up(u))-|c(u)\cap r(u)|),0\}$ and every other labels remain the same. 
\end{lem}

\begin{proof}
Let $L$ be a minimum $l(k,t)$-dominating set of $T$. Note that at any stage of the algorithm $k(up(u))=2$. Since $d(u)=2$, there exist a pendant edge $e$ incident to $u$ with $k(e)=2$. Therefore, $|(N_{T}[up(u)]\cup N_{T}[e])\cap L|\geq k(up(u))+k(e)-1$ which implies that $|(N_{T}[u]\cup N_{T}[p(u)])\cap L|\geq 3$. Since $k(e)\leq |r(u)|$, $r(u)$ contains at least two vertices from $N_T[u]$. This implies that $|r(u)\cup r(p(u))|\geq 2$. Since $|r(u)\cup r(p(u))|<3$, $(N_{T}[u]\cup N_{T}[p(u)])\cap L$ contains a set of vertices $Y$ of size at least one such that every vertex of $Y$ is of label $B$. Let $z$ be the maximum indexed vertex in $Y$ and $w$ be the maximum neighbour of $p(u)$ in $N_T[p(u)]\setminus r(p(u))$. Consider the set  $L'=(L\setminus \{z\})\cup \{w\}$. Clearly, $L'$ is a minimum $l(k,t)$-dominating set of $T$. Let $L''=L'\setminus (c(u)\cap r(u))$. Since $|N_{T}[up(u)]\cap L'|\geq k(up(u))$, we have $|N_{T'}[up(u)]\cap L''|\geq k'(up(u)) =\max\{(k(up(u))-|c(u)\cap r(u)|),0\}$. Since every other label remains same, for every other edge $e_j$ in $T'$, we have $|N_{T'}[e_j]\cap L''|\geq k'(e_j)$ and for every pair of edges $e_j$ and $e'_j$ in $T'$, we have $|(N_{T'}[e_j]\cup N_{T'}[e'_j])\cap L''|\geq k'(e_j)+k'(e'_j)-1$. Therefore, $L''$ is a $l(k',t')$-dominating set of $T'$. Hence, $\gamma_{l(k',t')}(T')\leq \gamma_{l(k,t)}(T)- |c(u)\cap r(u)|$.

Conversely, let $L'$ be a minimum $l(k',t')$-dominating set of $T'$. Let $w$ be the maximum neighbour of $p(u)$ in $N_T[u]\setminus r(u)$. Since $t'(w)=R$, we have $w\in L'$. Consider the set $L=L'\cup (c(u)\cap r(u))$. Since $L'$ is a $l(k',t')$-dominating set of $T'$, we have $|N_{T'}[up(u)]\cap L'|\geq k'(up(u))$. This implies that $|N_{T}[up(u)]\cap L|\geq k'(up(u))+|c(u)\cap r(u)|\geq k(up(u))$. Since every other label remains the same, for every edge $e$ which is not incident to $u$ in $T$, we have $|N_{T}[e]\cap L|\geq k(e)$ and for every pair of edges $e_j$ and $e'_j$ which are not incident to $u$ in $T$, we have $|(N_{T}[e_j]\cup N_{T}[e'_j])\cap L|\geq k(e_j)+k(e'_j)-1$. Since $k(e)\leq |r(u)|$ for every edge $e$ incident to $c(u)$, we have $|N_{T}[e]\cap L|\geq k(e)$ and $|r(u)|\geq 2$ because there is an edge $e'$ incident to a vertex of $c(u)$ with $k(e')=2$. Hence, we have $|N_T[up(u)]\cap L|\geq 3$. Therefore, the edge $up(u)$ satisfies condition three of $l(k,t)$-dominating set. Since $k(e)+k(e')-1\leq |r(u)|$, for every pair of distinct edges $e$ and $e'$ incident to the vertices of $c(u)$, we have $|(N_{T}[e]\cup N_{T}[e'])\cap L|\geq k(e)+k(e')-1$. Therefore, $L$ is a $l(k,t)$-dominating set of $T$. Hence, $\gamma_{l(k,t)}(T)\leq \gamma_{l(k',t')}(T')+|c(u)\cap r(u)|$. Therefore, $\gamma_{l(k,t)}(T)= \gamma_{l(k',t')}(T')+|c(u)\cap r(u)|$. \qed
\end{proof}

\begin{lem}\label{lem:d(u) neq 2}
Let $u$ be the support vertex at level $i$ processed in the second round. Also, assume that the labels of $T$ are such that $k(e)\leq |r(u)|$ for every edge $e$ incident to a vertex of $c(u)$ and $k(e)+k(e')-1\leq |r(u)|$ for every pair of distinct edges $e$ and $e'$ incident to vertices of $c(u)$. If $d(u)\neq 2$ or $|r(u)\cup r(p(u))|\geq 3$, then $\gamma_{l(k,t)}(T)=\gamma_{l(k',t')}(T')+|c(u)\cap r(u)|$, where $T'$ is obtained from $T$ by deleting the vertex set $c(u)$, setting  $k'(up(u))=\max\{(k(up(u))-|c(u)\cap r(u)|),0\}$ and every other labels remain the same. 	
\end{lem}

%	\noindent Proof of Lemma \ref{lem:d(u) neq 2}.
\begin{proof}
Let $L$ be a minimum $l(k,t)$-dominating set of $T$. Let $L'=L\setminus (c(u)\cap r(u))$. We show that $L'$ is a $l(k',t')$-dominating set of $T'$. Since $|N_T[up(u)]\cap L|\geq k(up(u))$, we have $|N_{T'}[up(u)]\cap L'|\geq k'(up(u))=\max\{(k(up(u))-|c(u)\cap r(u)|),0\}$. Since every other label remains same, for every other edge $e_j$, we have $|N_{T'}[e_j]\cap L'|\geq k'(e_j)$ and for every other pair of edges $e_j$ and $e'_j$, we have $|(N_{T'}[e_j]\cup N_{T'}[e'_j])\cap L'|\geq k'(e_j)+k'(e'_j)-1$. Therefore, $L'$ is a $l(k',t')$-dominating set of $T'$. Hence, $\gamma_{l(k',t')}(T')\leq \gamma_{l(k,t)}(T)- |c(u)\cap r(u)|$.

Conversely, let $L'$ be a minimum $l(k',t')$-dominating set of $T'$. Consider the set $L=L'\cup (c(u)\cap r(u))$. Since $L'$ is a $l(k',t')$-dominating set of $T'$, we have $|N_{T'}[up(u)]\cap L'|\geq k'(up(u))$. This implies that $|N_{T}[up(u)]\cap L|\geq k'(up(u))+|c(u)\cap r(u)|\geq k(up(u))$. Since every other label remains the same, for every edge $e_j$ which is not incident to $u$ in $T$, we have $|N_{T}[e_j]\cap L|\geq k(e_j)$ and for every pair of edges $e_j$ and $e'_j$ which are not incident to $u$, we have $|(N_{T}[e_j]\cup N_{T}[e'_j])\cap L|\geq k(e_j)+k(e'_j)-1$. Since $k(e)\leq |r(u)|$ for every edge $e$ incident to $c(u)$, we have $|N_{T}[e]\cap L|\geq k(e)$. Since $|k(e_j)+k(e'_j)-1|\leq r(u)$ for every pair of distinct edges $e_j$ and $e'_j$ incident to the vertices of $c(u)$, we have $|(N_{T}[e_j]\cup N_{T}[e'_j])\cap L|\geq k(e_j)+k(e'_j)-1$. Also, we have $|(N_T[up(u)]\cup N_T[e])\cap L|\geq k(up(u))+k(e)-1$ for every edge $e$ in $T$. Therefore, $L$ is a $l(k,t)$-dominating set of $T$. Hence, $\gamma_{l(k,t)}(T)\leq \gamma_{l(k',t')}(T')+|c(u)\cap r(u)|$. Therefore, $\gamma_{l(k,t)}(T)= \gamma_{l(k',t')}(T')+|c(u)\cap r(u)|$. \qed
\end{proof}

Finally, the following lemma shows how to compute a minimum $l(k,t)$-dominating set of a star $T$ rooted at $u$. 
\begin{lem}\label{Lem: u root du geq 2}
Let $u$ be the root of a star $T$. Then the following are true:
\begin{itemize}
\item[(a)] If $d(u)\geq 2$, then $\gamma_{l(k,t)}(T)=\max\{|r(u)|,3\}$.

\item[(b)] If $d(u)=1$, then $\gamma_{l(k,t)}(T)=\max\{|r(u)|,2\}$.

\item[(c)] If $d(u)=0$ and there is an edge $e$ incident to $u$ with $k(e)=1$, then $\gamma_{l(k,t)}(T)=\max\{|r(u)|,1\}$.

\item[(d)] If $d(u)=0$ and there is no edge $e$ incident to $u$ with $k(e)=1$, then $\gamma_{l(k,t)}(T)=|r(u)|$.
\end{itemize}	
\end{lem}

The proofs are straightforward and hence omitted. The above lemma shows how to compute minimum $l(k,t)$-dominating set of a given star.

\subsection{Running time of Algorithm \ref{Algo:treemain}}
In this subsection, we analyze the running time of Algorithm \ref{Algo:treemain}. Let $\sigma(i)$ denote the vertices at level $i$. The ordering $\sigma$ in line $1$ can be found in $O(n)$ time. While processing a support vertex $u$ at level $i$ in the for loop (line $4$--$11$), the algorithm marks some maximum neighbours of $u$ in $N_T[u]\setminus r(u)$ as $R$ and update the necessary $r()$-values. The conditions and the marking is done in constant time. Note that, a vertex can be marked as $R$ at most once. The update in line $11$ takes $O(deg_T(v_r))$ time. Therefore, the for loop in line $4$--$11$ takes $\sum\limits_{u\in \sigma(i)}\left(O(1)+O(deg_T(v_r))\right)$. With similar argument, the for loop in line $12$--$20$ takes $\sum\limits_{u\in \sigma(i)}\left(O(deg_T(u))+O(deg_T(v_r))\right)$ as line $16$-$18$ takes $O(deg_T(u))$ time. After summing for all the levels, we conclude that Algorithm \ref{Algo:treemain} runs in linear time.
Therefore, we have the following theorem:

\begin{theo}
The \textsc{MinLVEDP} can be solved in linear time for trees.
\end{theo}

\section{NP-completeness}

In this section, we show that the decision version of liar's ve-domination problem is NP-complete for general graphs. To prove the NP-completeness, we show a polynomial time reduction from liar's domination decision problem in general graphs, which is known to be NP-complete \cite{slater2009liar}. The decision version of liar's domination problem is stated as follows: 

\noindent\underline{\textsc{Liar's Domination Decision Problem} (\textsc{DecideLVDP})}

\noindent\emph{Instance}: A graph $G=(V,E)$ and an integer $k$.

\noindent\emph{Question}: Does there exist a liar's dominating set of $G$ of size at most $k$? 

\begin{theo}\label{Th:NPcompleteGen}
The \textsc{DecideLVEDP} is NP-complete for general graphs.
\end{theo}

\begin{proof}
Given a subset of the vertex set, we can check in polynomial time whether the given subset is a liar's ve-dominating set or not. Therefore, the \textsc{DecideLVEDP} is in NP.
Let $G=(V,E)$ be an instance of \textsc{DecideLVDP} with $V=\{v_1,v_2,\ldots,v_n\}$ and $E=\{e_1,e_2,\ldots,e_m\}$. We construct an instance of \textsc{DecideLVEDP}, say $G'=(V',E')$, by adding a vertex, say $u_i$, to every vertex $v_i\in V$. Therefore, $V'=V\cup\{u_i|1\leq i\leq n\}$ and $E'=E\cup\{u_iv_i|1\leq i\leq n\}$. Next, we prove the following lemma.

\begin{lem}\label{Lemma:LD=LVED}
The graph $G$ has a liar's dominating set of size at most $k$ if and only if $G'$ has a liar's ve-dominating set of size at most $k$.
\end{lem}

\begin{proof}
Let $L$ be a liar's dominating set of $G$ of size at most $k$. Thus, for every $v_i\in V$, we have $|N_{G}[v_i]\cap L|\geq 2$ and for any pair of vertices $v_i,v_j\in V$ with $i\neq j$, we have $|(N_{G}[v_i]\cup N_{G}[v_j])\cap L|\geq 3$. This implies that for every edge $v_iv_j$ and $u_iv_i$ in $G'$, we have $|N_{G'}[v_iv_j]\cap L|\geq 3$ and $|N_{G'}[u_iv_i]\cap L|\geq 2$, respectively. Also, for any pair of distinct edges $u_iv_i$ and $u_jv_j$, we have $|(N_{G'}[u_iv_i]\cup N_{G'}[u_jv_j])\cap L|\geq 3$. Similarly, we have $|(N_{G'}[u_iv_i]\cup N_{G'}[v_cv_j])\cap L|\geq 3$ for any pair of edges $u_iv_i$ and $v_cv_j$. Therefore, $L$ is a liar's ve-dominating set of $G'$ of size at most $k$.

Conversely, let $L_{ve}$ be a liar's ve-dominating set of $G'$ of size at most $k$. Note that to ve-dominate $u_iv_i$ twice, there must be at least two vertices from $N_{G'}[v_i]$ in $L_{ve}$. Let us assume that $u_i\in L_{ve}$ for some $i\in \{1,2,\ldots,n\}$. This implies that, $L_{ve}$ contains at least one vertex of $N_G[v_i]$. If $L_{ve}$ contains at least two vertices from $N_G[v_i]$ and at least three vertices from $(N_G[v_i]\cup N_G[v_j])$ for every $v_j\in V\setminus \{v_i\}$, then by removing $u_i$ from $L_{ve}$ we have another liar's ve-dominating set of $G'$ of size at most $k$. If $L_{ve}$ contains two vertices of $N_G[v_i]$ where $|N_{G}[v_i]|\geq 3$ and two vertices from $(N_{G}[v_i]\cup N_G[v_j])$ for some $v_j\in V\setminus \{v_i\}$, then replacing $u_i$ by a vertex $v_d\in N_{G}[v_i]\setminus L_{ve}$ we have another liar's ve-dominating set of $G'$ of size at most $k$. Also, if $L_{ve}$ contains two vertices from $N_G[v_i]$ where $|N_G[v_i]|=2$ and two vertices from $(N_{G}[v_i]\cup N_G[v_j])$ for some $v_j\in V\setminus \{v_i\}$, then replacing $u_i$ by a vertex $v_d\in N_G[v_j]\setminus L_{ve}$ we have a liar's ve-dominating set of $G'$ of size at most $k$. Finally, if $L_{ve}$ contains exactly one vertex from $N_G[v_i]$, then replacing $u_i$ by a vertex $v_d\in N_G[v_i]\setminus L_{ve}$ we have another liar's ve-dominating set of $G'$ of size at most $k$. Thus, without loss of generality, let us assume that $L_{ve}\subset V$. We claim that $L_{ve}$ is a liar's dominating set of $G$. If not, then there is a vertex $v_i\in V$ such that either $|N_G[v_i]\cap L_{ve}|<2$ or there is a vertex $v_j\in V\setminus \{v_i\}$ such that $|(N_G[v_i]\cup N_G[v_j])\cap L_{ve}|<3$. Since $L_{ve}$ is a liar's ve-dominating set of $G'$, to ve-dominate $u_iv_i$ twice, there must be at least two vertices of $N_G[v_i]$ in $L_{ve}$ and for any pair of edges $u_iv_i$ and $u_jv_j$ there must be at least three vertices of $N_G[v_i]\cup N_G[v_j]$ in $L_{ve}$, a contradiction. Therefore, $L_{ve}$ is a liar's dominating set of $G$ of size at most $k$. \qed
\end{proof}
From the above lemma, it follows that \textsc{DecideLVEDP} is NP-complete for general graphs. \qed 
\end{proof}

The \textsc{DecideLVDP} is NP-complete for chordal graph \cite{PANDAliarscomplexity} and bipartite graph \cite{roden2008liar}. Observe that if the instance $G=(V,E)$ of \textsc{DecideLVDP} is a chordal graph or bipartite graph, the graph $G'=(V',E')$ in Theorem \ref{Th:NPcompleteGen} is also chordal graph or bipartite graph, respectively. Thus, we have the following corollaries.
\begin{coro}
The \textsc{DecideLVEDP} is NP-complete for chordal graphs.
\end{coro}

\begin{coro}
The \textsc{DecideLVEDP} is NP-complete for bipartite graphs.
\end{coro}

The \textsc{DecideLVDP} is NP-complete in claw free graphs  \cite{panda2015hardness}.
\begin{coro}
The \textsc{DecideLVEDP} is NP-complete for $p$-claw free graphs where $p\geq 4$.
\end{coro}

\section{Approximation Algorithm and Hardness}
In this section, we study approximation algorithms and hardness of approximation for \textsc{MinLVEDP}.

\subsection{Approximation algorithms}\label{sec:Approxalgo}
In this subsection, we design two approximation algorithms for \textsc{MinLVEDP} problem in general graphs and in $p$-claw free graphs. 

\subsubsection{General graph}
Our proposed approximation algorithm uses the greedy approximation algorithm for the minimum set cover problem. The minimum set cover problem is stated as follows:

\noindent\underline{\textsc{Minimum Set Cover Problem} (\textsc{MinSCP})}

\noindent\emph{Instance}: A set system $(X,\mathcal{S})$, where $X$ is a nonempty set and $\mathcal{S}$ is a family of subsets of $X$.

\noindent\emph{Output}: A minimum set cover $\mathcal{C}$, that is, a subfamily $\mathcal{C}\subseteq \mathcal{S}$ such that every element $x$ of $X$ is contained in at least one of the sets in $\mathcal{C}$.
%$X=\bigcup\limits_{S\in \mathcal{C}} S$.

%\noindent\emph{Measure}: Cardinality of the set cover.

The approximation algorithm works in two phases. In the first phase, we compute a $2$-ve dominating set, say $D$, of the given graph $G=(V,E)$. For this we use the $(2\ln \Delta(G)+1)$-approximation algorithm by Kumar et al. \cite{naresh}, where $\Delta(G)$ is the maximum degree of $G$. Now if for every pair of edges $e_i, e_j\in E$, $|(N_G[e_i]\cup N_G[e_j])\cap D|\geq 3$, then $D$ is also liar's ve-dominating set of $G$. Otherwise, we construct an instance of the set cover problem as follows: let $X=\{(e_i,e_j)~such ~that~|(N_G[e_i]\cup N_G[e_j])\cap D|= 2\}$ and $\mathcal{S}$ be the family of subsets of $X$. For every vertex $v_k\in V\setminus D$, we have a set $S_k$ in $\mathcal{S}$, where $S_k$ is defined as $S_k=\{(e_{i},e_{j})\in X~such~ that~ v_k \in (N_{G}[e_{i}]\cup N_{G}[e_{j}])\}$. We find the approximate set cover, say $\mathcal{C}$, for the instance $(X, \mathcal{S})$ using the approximation algorithm in \cite{cormen}. Let $C$ be the set of vertices of $G$ defined by $C= \{v_k | S_k\in \mathcal{C}\}$. Note that $D$ and $C$ are disjoint subsets of $V$ and $L=D\cup C$ forms a liar's ve-dominating set of $G$. The process is described in Algorithm \ref{Algo:approximationgen}.

\begin{algorithm}
\caption{Approx\_LVEDS(G)}
\label{Algo:approximationgen}
\textbf{Input:} A graph $G=(V,E)$\\
\textbf{Output:} An approximate liar's ve-dominating set $L$.
\begin{algorithmic}[1]
\State Compute an approximate $2$-ve dominating set $D$;
\State Let $X=\{(e_i,e_j)~such ~that~|(N_G[e_i]\cup N_G[e_j])\cap D|= 2\}$
\State Let $S_k=\{(e_{i},e_{j})\in X~such~ that~ v_k \in (N_{G}[e_{i}]\cup N_{G}[e_{i}])\}$

\State $\mathcal{S}=\{S_k|v_k\in V\setminus D\}$;
\State Compute an approximate set cover $\mathcal{C}$ of $(X,\mathcal{S})$;
\State $C=\{v_k:S_k\in \mathcal{C}\}$;
\State $L=D\cup C$;
\State \Return $L$;

\end{algorithmic} 	
\end{algorithm}

%
%we consider the set of pair of edges $(e_i,e_j)$ for which $|(N[e_i]\cup N[e_j])\cap D|= 2$. Let $X=\{(e_i,e_j):|(N[e_i]\cup N[e_j])\cap D|= 2\}$. We find a set cover $C$ for the set $X$ and its family of subsets $S$ where each subset of $S$ has a neighbour in $V\setminus D$. The set $D\cup C$ is a liar's vertex-edge dominating set of $G$. The detail algorithm is given in Algorithm \ref{Algo:approximationgen}.

Next, we calculate the approximation ratio of the above algorithm. From \cite{naresh}, we know that $|D|\leq (2\ln \Delta(G)+1) |D^*|$, where $D^*$ is a minimum $2$-ve dominating set of $G$. As every liar's ve-dominating set is also a $2$-ve dominating set, we have $|D^*|\leq |L^*|$, where $L^*$ is a minimum liar's ve-dominating set of $G$. Also from \cite{cormen}, we have $|\mathcal{C}|\leq (\ln (|S_{max}|)+1) |\mathcal{C^*}|$, where $\mathcal{C^*}$ is a minimum set cover of $(X, \mathcal{S})$ and $S_{max}$ is the set in $\mathcal{S}$ having maximum cardinality. Clearly, $|\mathcal{C^*}|\leq |L^*|$ because corresponding sets of the vertices of $(L^*\setminus D)$ forms is a cover for $X$. Next, we argue about the cardinality of $S_{max}$. Let $S_{max}$ be the set corresponding to the vertex $v$. The number of edges that are exactly $2$-ve dominated by $D$ and also incident to a vertex of $N_G[v]$ is at most $\Delta(G)^2$. Let $e_i$ be such an edge and $N_{G}[e_i]\cap D= \{x,y\}$. Now, if $(e_i,e_j)\in X$, then $N_{G}[e_j]\cap D=\{x,y\}$. Note that $e_j$ must be an edge that is incident to $N_{G}[x]\cup N_{G}[y]$ and there are at most $\Delta(G)^2$ many edges possible. Hence, each $e_i$ contributes at most $\Delta(G)^2$ many pairs in $X$. As a result $S_{max}$ can contain at most $\Delta(G)^4$ many pairs from $X$. Therefore, $|S_{max}|\leq \Delta(G)^4$ and hence, $|\mathcal{C}|\leq (4\ln \Delta(G)+1) |\mathcal{C^*}|$. Combining all the above arguments, we have 
\begin{align*}
|L| = |D|+ |C| &= |D|+ |\mathcal{C}|\\
&\leq (2\ln \Delta(G)+1) |D^*| + (4\ln \Delta(G)+1) |\mathcal{C^*}|\\
&\leq (2\ln \Delta(G)+1) |L^*| + (4\ln \Delta(G)+1) |L^*|\\
&\leq 2(3\ln \Delta(G)+1) |L^*|
\end{align*}

Note that the algorithm can be computed in polynomial time. Hence, we have the following theorem:
\begin{theo}
The \textsc{MinLVEDP} problem can be approximated within a factor of $O(\ln \Delta(G))$, where $\Delta(G)$ is the maximum degree of $G$.
\end{theo}  

\subsubsection{$p$-claw free graph}\label{sec:p-claw}
In this subsection, we present a constant factor approximation algorithm for the \textsc{MinLVEDP} in $p$-claw free graphs. A graph $G=(V,E)$ is called a $p$-claw free if it does not have $K_{1,p}$ as an induced subgraph. A $3$-claw free graph is simply called a claw free graph. To this end, we first define a new parameter, namely \emph{distance $2$-matching} in a graph $G=(V,E)$. For a graph $G=(V,E)$, a distance $2$-matching is a subset of edges, say $M$, such that the distance between any two edges in $M$ is at least $2$. Here, the distance between a pair of edges $e_1=x_1y_1$ and $e_2=x_2y_2$ is defined as the length of a shortest path from $a$ to $b$ where $a\in \{x_1,y_1\}$ and $b\in \{x_2, y_2\}$. A distance $2$-matching $M$ is called maximal if $M$ is not contained in any other distance $2$-matching. Note that, a maximal distance $2$-matching can be computed greedily in linear time. In the following lemma, we relate the sizes of any maximal distance $2$-matching and any minimum $k$-ve dominating set for a $p$-claw free graph.

%We consider a maximal distance $2$-matching $M$ of $G$. Let $V(M)$ be the set of end points of the edges in the maximal distance $2$-matching $M$. The maximal distance $2$-matching can be found using a greedy algorithm.
\begin{lem}\label{Lem:p-clawfree}
Let $G=(V,E)$ be a $p$-claw free graph. Also let $D_k^*$ and $M$ be a minimum $k$-ve dominating set and a maximal distance $2$-matching of $G$, respectively. Then $\frac{k|M|}{(p-1)}\leq |D_k^*|$. 
\end{lem}

\begin{proof}
For every $e\in M$, let $c_e=|N_G[e]\cap D_k^*|$. Since $D_k^*$ is a $k$-ve dominating set of $G$, we have $c_e\geq k$. Therefore, we have 
\begin{equation}\label{Eq:ce}
\sum\limits_{e\in M}c_e\geq k|M|	
\end{equation}
For every vertex $v\in D_k^*$, let $d_v$ be the number of edges in $M$ which are ve-dominated by $v$. Since $G$ is a $p$-claw free graph, there are at most $p-1$ independent vertices in $N_G[v]$. Now, the distance between any two vertices in $V(M)$ must be at least $2$. Thus, for every vertex $v\in D_k^*$ we have $d_v\leq (p-1)$. This implies that, 
\begin{equation}\label{Eq:dv}
\sum\limits_{v\in D_k^*}d_v\leq (p-1)|D_k^*|	
\end{equation}
%	Now, for an edge $e\in M$, $c_e$ represents the number of vertices in $D_k^*$ which ve-dominates $e$. Also for a vertex $v\in D_k^*$, $d_v$ represents the number of edges in $M$ which are ve-dominated by $v\in D_k^*$. Thus, a vertex in $D_k^*$ contributes in $\sum\limits_{e\in M}c_e$ as many times as the number of edges of $M$ it ve-dominates. Also, a vertex $v\in D_k^*$ contributes in $\sum\limits_{v\in D_k^*}d_v$ at least as many times as the number of edges in $M$ which are ve-dominated by $v$.
Let us consider a matrix $A$ such that each row of $A$ corresponds to an edge $e_i$ in $M$ and each column of $A$ corresponds to a vertex $v_j$ in $D^*_k$. An entry $a_{ij}$ of $A$ is $1$ if the edge corresponding to $i$-th row is ve-dominated by the vertex corresponding to the $j$-th column and $0$ otherwise. Observe that, $c_e$ is the sum of the entries in a row of $A$ corresponding to the edge $e$. Thus, $\sum\limits_{e\in M}c_e$ is the sum of the entries of the matrix $A$. Also, $d_v$ is the sum of the entries in a column $A$ corresponding to the vertex $v$. Hence, $\sum\limits_{v\in D_k^*}d_v$ is the sum of the entries of the matrix $A$. 
Therefore, we have 
\begin{align*}
\sum\limits_{e\in M}c_e &= \sum\limits_{v\in D_k^*}d_v\\
k|M| &\leq (p-1)|D_k^*|\tag*{[Using inequality \ref{Eq:ce} and \ref{Eq:dv}]}\\
\frac{k|M|}{(p-1)} &\leq |D_k^*|
\end{align*} 
Hence, we have $\frac{k|M|}{(p-1)} \leq |D_k^*|$.  \qed
\end{proof}

Since every liar's ve-dominating set is also a $2$-ve dominating set, therefore we have the following corollary. 
\begin{coro}\label{Coro:p-clawfreeliar}
Let $G=(V,E)$ be any $p$-claw free graph. Also let $L^*$ and $M$ be a minimum liar's ve-dominating set and a maximal distance $2$-matching of $G$, respectively. Then $\frac{2|M|}{(p-1)}\leq |L^*|$.
\end{coro}

Based on Corollary \ref{Coro:p-clawfreeliar}, we propose the following approximation algorithm for \textsc{MinLVEDP} in $p$-claw free graphs. The outline of the algorithm is as follows: first, we compute a maximal distance $2$-matching $M_1$ in $G$. Then, we remove the edges $M_1$ from $G$ and compute another maximal distance $2$-matching $M_2$ in the remaining graph. Let $R$ be the set of edges that has an endpoint in $V(M_1)$ or in $V(M_2)$, where $V(M_i)$ is the set of endpoints of the edges in $M_i$. Next, we again remove the set of edges $R$ and compute another maximal distance $2$-matching say $M_3$ in the remaining graph. Finally, we return $L$ as a liar's ve-dominating set, where $L=L\cup (V(M_1)\cup V(M_2)\cup V(M_3))$. Since we can compute a maximal distance $2$-matching in linear time, clearly the algorithm runs in polynomial time. The process is described in Algorithm \ref{Algo:p-clawfree}.

\begin{algorithm}
\caption{Approx\_LVEDS\_$p$-ClawFree($G$)}
\label{Algo:p-clawfree}
\textbf{Input:} A $p$-claw free graph $G=(V,E)$.\\
\textbf{Output:} A liar's vertex-edge dominating set $L$.
\begin{algorithmic}[1]
\State $L=\emptyset$;
%		\State $M_0=\emptyset$;
%		\For{$(i=1~ to ~ 2)$}
\State Compute a maximal distance $2$-matching $M_1$ in $G$;
\State $G=G\setminus M_1$;
\State Compute a maximal distance $2$-matching $M_2$ in $G$;
%		\EndFor
\State $R=$ set of edges incident to $V(M_1)\cup V(M_2)$.
\State $G=G\setminus R$;
%		\State Remove all the edges from $E$ which is $3$-ve dominated by $L$.
\State Compute a maximal distance $2$-matching $M_3$ in $G$; 
\State $L=V(M_1)\cup V(M_2)\cup V(M_3)$;
%		\State $L=L\cup V(M_3)$;
\State \Return $L$;
\end{algorithmic}
\end{algorithm}
%Note that, Algorithm \ref{Algo:p-clawfree} is a polynomial time algorithm because, computing each $M_i$ for $i=1,2,3$ and removing the edges in step $6-7$ can also be done in polynomial time. 

Next, we prove that the set of vertices $L$, returned by Algorithm \ref{Algo:p-clawfree}, is a lair's ve-dominating set of $G$. Observe that, every edge $e\in E$ has at least one neighbour in $V(M_1)$. If possible, suppose that $e'$ be an edge that has no neighbour in $V(M_1)$. This implies that the distance between $e'$ and every edge in $M_1$ is at least $2$ which contradicts the fact that $M_1$ is a maximal distance $2$-matching. Therefore, $N_G[e]\cap V(M_1)\neq \emptyset$ for every edge $e\in E$. Also, every edge incident to $V(M_1)$ is $2$-ve dominated by $V(M_1)$. Now, by the similar argument as above every edge in $G\setminus M_1$ has a neighbour in $V(M_2)$. Clearly, $L$ $2$-ve dominates every edge in $E$. Also, for any pair of distinct edges in $M_1$ and $M_2$, the condition $2$ of liar's ve-domination holds. Now, suppose there is a pair of edges in $E\setminus (M_1\cup M_2)$ which are exactly $2$-ve dominated by the same pair of vertices in $V(M_1)\cup V(M_2)$. Every edge that is incident only to either $M_1$ or $M_2$ satisfies the condition $(2)$ of liar's ve-domination for $V(M_1)\cup V(M_2)$. So, the only edges which are exactly $2$-ve dominated by $V(M_1)\cup V(M_2)$ and do not satisfy condition $(2)$ of liar's ve-domination are the edges which have exactly one neighbour in $V(M_1)$ and exactly one neighbour in $V(M_2)$ but having no endpoint in $V(M_1)\cup V(M_2)$. Let $E''$ be the set of edges which are exactly $2$-ve dominated by $V(M_1)\cup V(M_2)$. Clearly, $E''$ is a subset of the set of edges of $G\setminus R$. Now, each edge in $E''$ has a neighbour in $V(M_3)$ otherwise there must be an edge in $E''$ which is at distance $2$ from every edge in $M_3$ but is not in $M_3$ contradicting the maximality of $M_3$.  
Thus, every edge $e_i$ in $G$ is $2$-ve dominated by $L$ and for every pair of edges $e_i$, $e_j$ in $G$, we have $|(N_G[e_i]\cup N_G[e_j])\cap L|\geq 3$. Therefore, $L=V(M_1)\cup V(M_2)\cup V(M_3)$ is a liar's ve-dominating set of $G$.

Finally, we calculate the approximation ratio of the algorithm described above. From Corollary \ref{Coro:p-clawfreeliar}, we have $2 |M_i|\leq (p-1)|L^*|$ for every $1 \leq i\leq 3$. Thus, we have
\begin{align*}
|L| & \leq |V(M_1)| + |V(M_2)| + |V(M_3)|\\
&\leq 2|M_1|+2|M_2|+2|M_3|\tag*{[Since $|V(M_i)|=2|M_i|$, for every $1\leq i\leq 3$]}\\
&\leq 3(p-1)|L^*|
\end{align*}
Therefore, we have the following theorem:
\begin{theo}
The \textsc{MinLVEDP} can be approximated within a factor of $3(p-1)$ in $p$-claw free graphs.
\end{theo}

\subsection{Lower bound on approximation ratio}\label{Approxhard}

In this subsection, we show a lower bound on the approximation ratio of \textsc{MinLVEDP} by reducing instances of minimum liar's domination problem into instances of this problem. The minimum liar's domination problem is defined as follows: 
%The problem is described as follows-
%
%\noindent\underline{\textbf{Minimum liar's vertex-edge domination Problem}(\textsc{MinLVEDP})}
%
%\noindent\emph{Instance}: A graph $G=(V,E)$.
%
%\noindent\emph{Solution}: A liar's vertex-edge dominating set $L_{ve}$ of $G$.
%
%\noindent\emph{Measure}: Cardinality of the liar's vertex-edge dominating set.
%To prove the lower bound we show a reduction from liar's dominating set problem. The problem is as follows-

\noindent\underline{\textbf{Minimum liar's domination Problem}(\textsc{MinLDP})}

\noindent\emph{Instance}: A graph $G=(V,E)$.

\noindent\emph{Output}: A minimum liar's dominating set $L$ of $G$.

%\noindent\emph{Measure}: Cardinality of the liar's dominating set.

Panda et al. \cite{panda2015hardness} showed that given a graph $G=(V,E)$, the \textsc{MinLDP} cannot be approximated within $(\frac{1}{8}-\epsilon)\ln|V|$ for any $\epsilon >0$, unless $NP\subseteq DTIME(|V|^{O(\log(\log|V|)})$.
%\begin{theorem}[\cite{panda2015hardness}]
%	For a graph $G=(V,E)$, the \textsc{MinLDP} can not be approximated within $(\frac{1}{8}-\epsilon)\ln|V|$ for any $\epsilon >0$, unless $NP\subseteq DTIME(|V|^{O(\log(\log|V|)})$.
%\end{theorem}
Using this result, we prove a lower bound on the approximation ratio by showing an approximation preserving reduction from \textsc{MinLDP} to \textsc{MinLVEDP}.
%Next, by using the above construction, we show the lower bound on the approximation ratio.
We use the same construction as described in Theorem \ref{Th:NPcompleteGen}. In the next theorem, we show the lower bound on the approximation ratio for general graphs.
\begin{theo}
For a graph $G=(V,E)$, the \textsc{MinLVEDP} cannot be approximated within a factor of $\frac{1}{2}(\frac{1}{8}-\epsilon)\ln|V|$ for any $\epsilon >0$, unless \\$NP\subseteq DTIME(|V|^{O(\log(\log|V|)})$.
\end{theo}
\begin{proof}
We prove this by contradiction. If possible, let $\mathcal{A}$ be an $\alpha$-approximation algorithm to compute minimum liar's ve-dominating set of $G'=(V',E')$, where $\alpha=\frac{1}{2}(\frac{1}{8}-\epsilon)\ln|V'|$. We can compute the minimum liar's dominating set of a given graph $G=(V,E)$ as follows: given a graph $G=(V,E)$ we first construct the graph $G'$ from $G$ as described in Theorem \ref{Th:NPcompleteGen}. Then compute an approximate liar's ve-dominating set $L_{ve}$ of $G'$ using algorithm $\mathcal{A}$. By Lemma \ref{Lemma:LD=LVED}, we have a liar's dominating set, say $L$, of $G$ such that $|L|\leq |L_{ve}|$. Clearly, this can be computed in polynomial time. Let $L^*$ and $L^*_{ve}$ be the minimum liar's dominating set of $G$ and minimum liar's ve-dominating set of $G'$, respectively.
Therefore, we have
\begin{align*}
|L| \leq |L_{ve}| &\leq \frac{1}{2}(\frac{1}{8}-\epsilon)\ln|V'| |L^*_{ve}|\\
&\leq (\frac{1}{8}-\epsilon)\ln|V| |L^*_{ve}| \tag*{$[Since~ |V'|=2|V|\leq |V|^2]$}\\
&=  (\frac{1}{8}-\epsilon)\ln|V| |L^*| \tag*{$[Since~ |L^*|= |L^*_{ve}|~by~ Lemma~ \ref{Lemma:LD=LVED}]$}
\end{align*}
Therefore, the above process actually outputs an approximate liar's dominating set of $G=(V,E)$ with approximation ratio $(\frac{1}{8}-\epsilon)\ln|V|$. This is a contradiction. Hence, the \textsc{MinLVEDP} cannot be approximated within $\frac{1}{2}(\frac{1}{8}-\epsilon)\ln|V|$ for any $\epsilon >0$, unless $NP\subseteq DTIME(|V|^{O(\log(\log|V|)})$. \qed
\end{proof}
\subsection{APX-completenss}

In \cite{panda2015hardness}, we have that the \textsc{MinLVDP} is APX-complete for graphs with maximum degree $4$ and $p$-claw free graphs for $p\geq 5$. Observe that the reduction in Theorem \ref{Th:NPcompleteGen} is a $L$-reduction. Algorithm \ref{Algo:approximationgen} and Algorithm \ref{Algo:p-clawfree} are constant factor approximation algorithms for bounded degree graphs and $p$-claw free graphs for $p\geq 4$ respectively. Therefore, we have the following theorem. 
\begin{theo}
The \textsc{MinLVEDP} is APX-complete for graphs with maximum degree $5$.
\end{theo}  

\begin{theo}
The \textsc{MinLVEDP} is APX-complete for $p$-claw free graph for $p\geq 6$.
\end{theo}

\section{Conclusion}
In this paper, we have introduced the notion of liar's ve-domination motivated by an application. Our study mainly focuses on algorithmic aspects of this problem. We have shown that \textsc{MinLVEDP} is solvable in linear time for trees but it is NP-complete for chordal graphs, bipartite graphs, and $p$-claw free graphs for $p\geq 4$. We have also shown that \textsc{MinLVEDP} can be approximated with approximation ratio $O(\ln \Delta(G))$, where $\Delta(G)$ is the maximum degree of the input graph. But \textsc{MinLVEDP} cannot be approximated within $\frac{1}{2}(\frac{1}{8}-\epsilon)\ln|V|$ for any $\epsilon >0$, unless $NP\subseteq DTIME(|V|^{O(\log(\log|V|)})$. Also, we have presented a $3(p-1)$-approximation algorithm for $p$-claw free graphs. Further, we have shown that the \textsc{MinLVEDP} is APX-complete for bounded degree graphs and $p$-claw free graphs for $p\geq 6$. It would be interesting to study the status of this problem in subclasses of chordal and bipartite graphs. Also, designing better approximation algorithms for other graph classes can be studied.

\bibliographystyle{alpha}     % mathematics and physical sciences
\bibliography{VEDom_bib}   % name your BibTeX data base

\newcommand{\etalchar}[1]{$^{#1}$}
\begin{thebibliography}{NKPV21}

\bibitem[ACS{\etalchar{+}}21]{ahangar2021total}
H~Abdollahzadeh Ahangar, M~Chellali, SM~Sheikholeslami, M~Soroudi, and
  L~Volkmann.
\newblock Total vertex-edge domination in trees.
\newblock {\em Acta Mathematica Universitatis Comenianae}, 90(2):127--143,
  2021.

\bibitem[BC18]{Totalve-domchellali}
Razika Boutrig and Mustapha Chellali.
\newblock Total vertex-edge domination.
\newblock {\em International Journal of Computer Mathematics},
  95(9):1820--1828, 2018.

\bibitem[BCHH16]{boutrig2016vertex}
R.~Boutrig, M.~Chellali, T.~W Haynes, and S.T. Hedetniemi.
\newblock Vertex-edge domination in graphs.
\newblock {\em Aequationes mathematicae}, 90:355--366, 2016.

\bibitem[CLRS01]{cormen}
Thomas~H. Cormen, C.~E. Leiserson, R.~L. Rivest, and C.~Stein.
\newblock {\em Introduction to algorithms,second edition}.
\newblock The MIT Press, Massachusetts, 2001.

\bibitem[CS12a]{chitra2012global}
S.~Chitra and R.~Sattanathan.
\newblock Global vertex-edge domination sets in graph.
\newblock In {\em International Mathematical Forum}, volume~7, pages 233--240,
  2012.

\bibitem[CS12b]{globalvedom}
S.~Chitra and R.~Sattanathan.
\newblock Global vertex-edge domination sets in total graph and product graph
  of path {$P_n$} cycle {$C_n$}.
\newblock In {\em Mathematical modelling and scientific computation}, volume
  283 of {\em Commun. Comput. Inf. Sci.}, pages 68--77. Springer, Heidelberg,
  2012.

\bibitem[JD22]{jena}
S.K. Jena and G.K. Das.
\newblock Vertex-edge domination in unit disk graphs.
\newblock {\em Discrete Applied Mathematics}, 319:351--361, 2022.

\bibitem[JJD20]{unitdiscliar}
Ramesh~K. Jallu, Sangram~K. Jena, and Gautam~K. Das.
\newblock Liar's domination in unit disk graphs.
\newblock {\em Theoretical Computer Science}, 845:38--49, 2020.

\bibitem[KCV17]{krishna}
B.~Krishnakumari, M.~Chellali, and Y.B. Venkatakrishnan.
\newblock Double vertex-edge domination.
\newblock {\em Discrete Mathematics, Algorithms and Applications},
  9(04):1750045, 2017.

\bibitem[Lew07]{lewis}
J.R. Lewis.
\newblock {\em Vertex-edge and edge-vertex parameters in graphs.}
\newblock PhD thesis, Clemson University, Clemson, SC, USA, 2007.

\bibitem[LW23]{li2023polynomial}
Peng Li and Aifa Wang.
\newblock Polynomial time algorithm for k-vertex-edge dominating problem in
  interval graphs.
\newblock {\em Journal of Combinatorial Optimization}, 45(1):45, 2023.

\bibitem[NKPV21]{naresh}
H.~Naresh~Kumar, D.~Pradhan, and Y.B. Venkatakrishnan.
\newblock Double vertex-edge domination in graphs: complexity and algorithms.
\newblock {\em Journal of Applied Mathematics and Computing}, 66(1):245--262,
  2021.

\bibitem[Pet86]{peters}
K.W.J. Peters.
\newblock {\em Theoritical and algorithmic results on domination and
  connectivity}.
\newblock PhD thesis, Clemson University, Clemson, SC, USA, 1986.

\bibitem[PP13]{PANDAliarscomplexity}
B.S. Panda and S.~Paul.
\newblock Liar’s domination in graphs: Complexity and algorithm.
\newblock {\em Discrete Applied Mathematics}, 161(7):1085--1092, 2013.

\bibitem[PPP15]{panda2015hardness}
Bhawani~Sankar Panda, Subhabrata Paul, and Dinabandhu Pradhan.
\newblock Hardness results, approximation and exact algorithms for liar's
  domination problem in graphs.
\newblock {\em Theoretical Computer Science}, 573:26--42, 2015.

\bibitem[PPV21]{paul2}
S.~Paul, D.~Pradhan, and S.~Verma.
\newblock Vertex-edge domination in interval and bipartite permutation graphs.
\newblock {\em Discussiones Mathematicae: Graph Theory}, 43(4):947--963, 2021.

\bibitem[PR22]{paul}
S.~Paul and K.~Ranjan.
\newblock Results on vertex-edge and independent vertex-edge domination.
\newblock {\em Journal of Combinatorial Optimization}, 44(1):303--330, 2022.

\bibitem[RS09]{roden2008liar}
Miranda~L. Roden and Peter~J. Slater.
\newblock Liar’s domination in graphs.
\newblock {\em Discrete Mathematics}, 309(19):5884--5890, 2009.

\bibitem[Sla09]{slater2009liar}
Peter~J Slater.
\newblock Liar's domination.
\newblock {\em Networks: An International Journal}, 54(2):70--74, 2009.

\bibitem[{\.Z}yl19]{zylinski2019vertex}
P.~{\.Z}yli{\'n}ski.
\newblock Vertex-edge domination in graphs.
\newblock {\em Aequationes Mathematicae}, 93(4):735--742, 2019.

\end{thebibliography}
%
%% Non-BibTeX users please use
%\begin{thebibliography}{}
%%
%% and use \bibitem to create references. Consult the Instructions
%% for authors for reference list style.
%\end{thebibliography}

\end{document}